\documentclass[preprint, prl, showpacs,preprintnumbers,
amsmath,amssymb]{revtex4}
\usepackage{graphicx}

\newcommand{\bc}{\ensuremath{\mathbf{c}}}
\newcommand{\bu}{\ensuremath{\mathbf{u}}}
\newcommand{\bx}{\ensuremath{\mathbf{x}}}

\newcommand{\LagV}{\ensuremath{\boldsymbol{B}}}

\newcommand{\bnabla}{\ensuremath{\boldsymbol{\nabla}}}
\begin{document}
\title
{Minimal entropic kinetic  models for simulating hydrodynamics}
\author{Santosh Ansumali}
\author{ Iliya V.\ Karlin}
\thanks{Corresponding author. E-mail: ikarlin@ifp.mat.ethz.ch}
 \author{Hans Christian \"Ottinger}
\affiliation{ETH-Z\"urich, Department of Materials, Institute of Polymers\\
ETH-Zentrum, Sonneggstr. 3, ML J 19, CH-8092 Z\"urich, Switzerland}
\begin{abstract}

We derive minimal  discrete models of the Boltzmann equation
consistent with equilibrium thermodynamics, and which recover correct hydrodynamics in arbitrary
dimensions. 
A simple analytical procedure of constructing the equilibrium for the nonisothermal 
hydrodynamics is established.  A new discrete velocity model is
proposed for the simulation of the Navier-Stokes-Fourier equation and
is tested in the set up of Taylor vortex flow. For the  lattice Boltzmann method of  isothermal hydrodynamics, 
the explicit analytical form of the equilibrium distribution is presented. 

\end{abstract}
\maketitle
 Minimal kinetic models, and primarily the lattice Boltzmann
method  (hereafter LBM), have recently met with significant success in simulations of complex
isothermal hydrodynamic phenomena. A few examples of the
 successful application are the simulation of fluid-particle
 suspensions, turbulent flows, spinodal decomposition   \cite{succi,
  Rev, Cates}. The first large-scale simulations of 3D spinodal
decomposition in inertial regime \cite{Cates} and simulation of
Brownian short-time regime
\cite{LaddST} are a few successes achieved in the field of
computational fluid mechanics through use of these approaches.  The ability to
handle a very complicated geometry in a very simple manner has allowed
these method to emerge as an alternative to purely continuum
approaches, even for solving isothermal Navier-Stokes equation \cite{Rev}. In these methods, hydrodynamic equations
(for example, the isothermal Navier-Stokes and the nonisothermal Navier-Stokes-Fourier equations in our terminology) are 
not addressed by a direct discretization procedure. Instead a simplified  kinetic equation is introduced in such a way that hydrodynamic equations are obtained as its
 large-scale long-time limit. Two central issues in the construction of such models are 
the choice of discrete velocities, as few as possible, and the construction of the local equilibria such that the
 desired hydrodynamic equations are reproduced as closely as
 possible by the kinetic model. 

 Unlike the isothermal case, the kinetic modeling of the nonisothermal
 hydrodynamics is a hitherto unsolved problem \cite{RMP, succi}. The
 kinetic models with proper thermodynamics are especially needed for
 the simulation of 
  chemically reactive flows and the multiphase flows and near
 continuum flows in microdevices, which are
 difficult to simulate using a purely continuum models.  
 However,
 the nonisothermal model is not established even in the case of the
 Navier-Stokes-Fourier dynamics of a single fluid. 
  Apart from the conserved moments of distribution function, the mass, the  momentum and
   the energy, also non-conserved moments, the stress tensor, the heat
 flux and  fourth moments need to be
   in a specific form to recover the Navier-Stokes-Fourier equations. 
 This
  was previously accomplished  by assuming a pre-defined simple
 functional form, typically a polynomial, for the equilibrium population \cite{succi}.
In such a setting, the construction is neither unique in the choice of
  the discrete velocity set, nor in  the choice of the
  function. Moreover, these schemes permit populations to attain
  negative values and thus make the simulation scheme unstable
  \cite{RMP,Alder}. The way to resolve the problem of  non-positive
  form of the  population is to define the
  equilibrium population as a
  minimum of a convex function, known as $H$ function,  under the constraint of local
  conservation laws. Recently, the advantage of such an approach was
  shown in the context of  two dimensional isothermal hydrodynamics
 \cite{DHT, AK1, AK2,AK3}. 
 In order to avoid an explicit calculation of the local
  equilibria, these studies took an alternative root of using 
  computationally intensive kinetic equations.

 Apart from the stability issue, another well known
  problem associated with the current discrete velocity models is
  non-adherence to the equation of the state \cite{DTD}. 
In these models,  the local
  equilibrium entropy does not  satisfy the usual
definition of the temperature as a function of  the entropy  and the
  energy known from the elementary thermodynamics \cite{DTD}.

In this letter, we construct   kinetic models, which are free from  all
the problems discussed above,  in arbitrary
dimensions. The proper choice of the set of the  discrete velocities
and the $H$ function  and the explicit expression for the equilibrium
are the main result of the present work. 
 These models retain the simplicity and
computational efficiency of the standard lattice Boltzmann model.
 Further, for the  isothermal lattice Boltzmann  method
 an explicit equilibria with correct $H$ theorem is derived.

  Before constructing such model, we briefly explain  the  basic setup of the discrete
 velocity models. In these methods, the kinetic equation is written
 for the populations $f_i(\bx,t)$ of the discrete velocities 
 $\bc_i$, $i=1,\dots,b$, defined at position $\bx$ and time $t$. 
Hydrodynamic fields are first few  moments of populations, namely
$\rho = \sum_{i=1}^b f_i$ (density), 
$\rho u_{\alpha}  = \sum_{i=1}^b f_i \mathbf{c}_{i\alpha}$ (momentum density,
$\alpha=1,\dots, D$, where $D$ is the spatial dimension), and 
$\rho D T+ \rho u^2 = \sum_{i=1}^b f_i \mathbf{c}_i^2$ (energy density).
In the case of isothermal hydrodynamics, the 
hydrodynamic fields include $\rho$ and $\rho u_{\alpha}$, 
whereas in the nonisothermal case the energy density
is also included as  independent field.
Typical model kinetic equation reads,
\begin{equation}
\label{LBM}
\partial_t f_i+ c_{i\alpha} \partial_{\alpha}f_i =
-\tau^{-1}
\left(f_i- f_i^{\rm eq}  \right),
\end{equation}
where the model collision integral on the right hand side is assumed in the
 Bhatnagar-Gross-Krook (BGK) form
\cite{Cerci}, with  $\tau$ as the relaxation time.
The collision integral must respect local conservation laws which imply (in the nonisothermal case),
\begin{equation}
\label{mom}
\sum_{i=1}^b f^{\rm eq}_i \{ 1,\  c_{i\alpha},\ c_i^2 \}=\{\rho,\ \rho u_{\alpha},\ \rho DT+ \rho u^2 \}
\end{equation}
In the isothermal case the energy constraint on the local equilibrium  is excluded from this list.
Besides Eq.\ (\ref{mom}), the local equilibrium must respect several
other conditions for the non-conserved fields (examples will be given below).
These latter 
conditions are found on the basis of  the Chapman-Enskog 
analysis of the model (\ref{LBM}), and they ensure that the 
long-time large-scale dynamics (\ref{LBM}) would be the desired hydrodynamic equations.
The problem is then  reduced to finding  a parametric expression for the
equilibrium population such  that all the constraints are satisfied.

However, in order to construct minimal entropic model we take a different root.  
To begin with, we remind the reader of the important observation \cite{DLBM}
on the relation between the discretization of the velocity space and the well known
Grad's moment method  \cite{Grad}. Namely, if 
discrete velocities are constructed from  zeros of the Hermite polynomials,
the method of
discrete velocity is essentially equivalent to  Grad's  moment
method  based on the expansion of the distribution
function around a fixed Maxwellian distribution function.
The natural extension of this approach towards entropic schemes is to link the 
discrete velocity model not to the Grad's method but instead to the entropic Grad's method
(the maximum entropy approximation) \cite{Kogan65}.  
To this end, Boltzmann's $H$ function, $H=\int F\ln Fd\bc$, where
$F(\mathbf{x}, \bc)$ is the one-particle distribution function, 
$\mathbf{x}$ is the  position vector, and
$\bc$ is the continuous  velocity, is 
evaluated  using the Gauss-Hermite quadrature. 
This  gives the discrete form of the $H$ function,

\begin{equation}
\label{app:H}
H_{\{w_i,\bc_i\}}=\sum_{i=1}^{b} f_{i}\ln\left(\frac{f_{i}}{w_i} \right).
\end{equation}
Here $w_i$ is the weight associated with the $i$th discrete velocity
$\bc_i$, and the  
particles mass and the Boltzmann constant $k_{\rm B}$ are set equal to
the unity. 
Discrete velocity populations $f_i(\mathbf{x})$ are related to 
values of the distribution function at the nodes of the quadrature as
 $f_i(\mathbf{x})=w_i(2\, \pi \, T_0)^{(D/2)} \exp(c^2_i/(2\,T_0))F(\mathbf{x},
 \mathbf{c}_i)$. Note that the weights are incorporated into the
 definition  of $f_i$, and that the Maxwell velocity distribution function has
 been factored out because this Gaussian  probability distribution is
 taken into account through the Gauss--Hermite quadrature.   
Discrete-velocity entropy functions (\ref{app:H}) for various $\{w_i,\bc_i\}$ is the
unique input for all our constructions below.

We shall first consider the isothermal
hydrodynamics. It is
known \cite{DLBM} that the minimal set of discrete velocities needed to reconstruct
Navier-Stokes equations corresponds to  zeroes of the third order
Hermite polynomials in $c_ i/ \sqrt{2 \, T_0}$. 
For $D=1$, the three discrete velocities are
$ c = \{-\sqrt{3\,T_0}, 0, \sqrt{3\,T_0}\}$,  whereas the corresponding weights are
 $w = \left \{\frac{1}{6},\frac{2}{3}, \frac{1}{6} \right \}$ respectively. 
In higher dimensions, the discrete
velocities are tensor products of the discrete velocities in one
dimension  and the weights are constructed by multiplying weights
associated with each component direction. 
In this way we construct the entropy function (\ref{app:H}). It is important to remark that 
for $D=2$
the entropy
function thus obtained coincides with the
one derived in  Ref.\ \cite{DHT}  by a completely
different kind of argument.

The discrete-velocity local equilibrium is the minimizer of the corresponding 
entropy function under the fixed density and the momentum (\ref{mom}). 
The explicit solution to this conditional minimization problem in $D$
dimensions reads:

 \begin{equation}
\label{TED}
 f^{\rm eq}_i=\rho w_i\prod_{\alpha=1}^{D} 
\left(2 -\sqrt{1+ 
{u_{\alpha}^{\prime}}^2}
\right)
\left(
\frac{\frac{2}{\sqrt{3}}u_{\alpha}^{\prime}+ 
\sqrt{1+ {u_{\alpha}^{\prime}}^2}}{1-u_\alpha^{\prime}/(\sqrt{3})}
\right)^{c_{i\alpha}/\sqrt{3}c_{\rm s}},
\end{equation}

where $c_{\rm s}=\sqrt{\,T_0}$ is the speed of the sound and the dimensionless
velocity $u_{\alpha}^{\prime}=\left({u_\alpha}/{c_{\rm s}}\right)$. Note that the
exponent, $(c_{i\alpha}/(\sqrt{3}c_{\rm s}))$, in Eq. (\ref{TED}) takes
 the values $\pm 1, \, \mbox{and}\,\,  0$ only and the resulting
 expressions for the equilibrium can be simplified in each dimensions. 
Equilibria (\ref{TED}) are positive definite for $u_{\alpha}<\sqrt{3}c_{\rm s}$. 
Without going into details of derivation, we mention that the equilibrium (\ref{TED}) 
is the product of $D$ one-dimensional solutions, see for example Ref.\ \cite{AK1}. This factorization is pertinent to the derivation,
and is quite similar to the familiar  property of Maxwell's distribution function.

 We stress that none of the conditions for the higher-order equilibrium moments
have been used while deriving (\ref{TED}). Relevant  higher-order moments of the equilibrium distribution, needed to establish
isothermal hydrodynamics in the framework of the Chapman-Enskog method  are  
the equilibrium pressure tensor,
$P_{\alpha \beta}^{\rm eq}=\sum_{i}f_i^{\rm eq}c_{i\alpha}c_{i\beta}$,  
and the equilibrium third-order moments,
$Q_{\alpha \beta \gamma }^{\rm eq}=\sum_{i}f_i^{\rm eq}c_{i\alpha}c_{i\beta}c_{i\gamma}$. 
A direct computation shows that the present equilibrium results in the correct
pressure tensor (correct means as obtained in the continuous kinetic
theory) to the order $O(u^4)$, which is sufficient to simulations of
the Navier-Stokes dynamics at small Mach
number. For example, for $D=2$, even at high dimensionless velocity, 
$u_\alpha = 0.25\sqrt{\,T_0}$, the error in the  pressure is less than
$2\%$.
Moreover, the {\it third-order} moment is even {\it more accurate}
as compared to the standard quadratic polynomial equilibrium, which 
leads to a $O(u^3)$ error in simulations
 \cite{Orszag}. In the present case,   only the diagonal component $Q^{\rm eq}_{\alpha\alpha\alpha}$
have the same linear accuracy as in the standard 
LBM, whereas all the  $D^3 -D$ non-diagonal components of 
 $Q_{\alpha \beta \gamma}^{\rm eq}$ are correct up to  order
 $O(u^5)$.

      The set of the discrete velocities used in the present case is the
      same as standard $D2Q9$ and $D3Q27$ model of the lattice Boltzmann
      method. The expansion of the equilibrium to the order $O(u^2)$
      coincides with the polynomial equilibria used in the lattice
      Boltzmann method. Thus, it is not surprising that $D2Q9$ model
      is more stable than any other LBM.
   However, due to the enhanced numerical stability and accuracy in
      the heat flux, the use of the  positive definite equilibrium
      (Eq. \ref{TED}) is more preferable in
      comparison to its expanded form in  the isothermal lattice BGK method.

In precisely the same way, the minimal entropic kinetic model for the nonisothermal case requires 
zeroes of fourth-order  Hermite polynomials. 
For $D=1$, the four discrete velocities and corresponding weights
of Gauss-Hermite quadrature  are $c=\left (\pm a, \pm b \right)$, and 
$w=(w_a, w_b)=\left(T_0/(4a^2) , T_0/(4b^2) \right)$,  respectively,
where $a=\sqrt{3-\sqrt{6}}(\,T_0)^{1/2}$, and $b=\sqrt{3+\sqrt{6}}(\,T_0)^{1/2}$.
The  minimizer of the $H$ function (\ref{app:H}) 
corresponding to the velocity set and weights just described,
and subject to the constraints (\ref{mom}), may be written as

\begin{align}
\label{1Dthermo}
f^{\rm eq}_i=\begin{cases}
 \frac{\rho\left(b^2 - T - u^2\right)}{b^2 - a^2}
\frac{ \exp{( B\, c_i)}}{\exp{(B \,a)}+\exp{(- B \, a)}} &
 \text{if $c_i = \pm a$},\\
\frac{\rho\left(u^2 + T - a^2\right)}{b^2 - a^2}
\frac{ \exp{( B\, c_i)}}{\exp{(B \,b )}+\exp{(- B \, b)}} &
 \text{if $c_i = \pm b$}
\end{cases}
\end{align}

Lagrange multiplier $B$, corresponding to the momentum constraint, has the following 
series representation:
\begin{equation}
\label{1DB}
 B= \frac{u}{E} -
 \frac{ u^3}{3}\biggl [ \frac{ a^2 b^2}{E^4} -
 \frac{( a^2 + b^2)}{E^3}\biggr ] 
 + O(u^5E^{-5}),
\end{equation}
where $E=u^2+T$ is the total energy density 
[Notice that Eq.\ (\ref{1DB})  is not an expansion in powers of velocity $u$, rather, in terms of 
$u^nE^{-m}$.]
Equilibrium distribution (\ref{1Dthermo}) exists within a positivity
interval, $ a^2 \le T + u^2 \le b^2$.

In higher dimensions, the set of discrete velocities is formed  by
taking tensor product of the discrete velocities in $D=1$, and the weights are products of
corresponding quantities.  In order to evaluate Lagrange multipliers in the 
formal solution to the minimization problem,  
$f^{\rm eq}_i=w_i \exp{\left( A +  \LagV \cdot\bc_i + Cc_i^2\right)}$, we first make an important 
observation
that they can be computed exactly for $\bu=0$ and {\it any} temperature $T$ within the
positivity interval, $a^2<T<b^2$:

\begin{align}
\begin{split}
B_\alpha = 0, \; \;
C_0 =\frac{1}{(b^2 - a^2)} \log { \left( \frac{w_a\, (T-a^2)}{w_b\, (
      b^2 - T)}\right)},
\\
A_0 =  \log{\left( \frac{\rho\, (b^2 -T)^D }{ (2 w_a)^D ( b^2 -a^2)^D}\right)} -
D\,a^2 C_0. 
\end{split}
\end{align}

With this, we find the equilibrium  at zero average velocity and
arbitrary temperature,

\begin{eqnarray}
\label{TH0}
f^{\rm eq}_i
 &=& 
\frac{\rho\, w_i}{2^D(b^2 - a^2)^D}\times
\\\nonumber
&&
\prod_{\alpha=1}^{D}
\left(\frac{b^2 - T  }{w_a} \right)^{
\left(\frac{b^2 - c_{i \alpha}^2 }{b^2 - a^2} \right)}
\left(\frac{T - a^2 }{w_b} \right)^{
\left(\frac{c_{i \alpha}^2 - a^2 }{b^2 - a^2} \right)}
. 
\end{eqnarray}
Factorization over spatial components is clearly seen in this solution.
Once the exact solution for zero velocity  is known, extension to $\bu\ne0$
is easily found by perturbation. The first few terms of this expansion are: 
\begin{eqnarray*}
A &=& A_0-\frac{T}{( T - a^2)(b^2 - T)}u^2+ O(u^4),
\\
B_{\alpha}&=&\frac{u_\alpha}{ T}  + \frac{(T - T_{0})^2}{2D T^4} 
\left(D\, u_\beta u_\theta u_\gamma \delta_{\alpha \beta \gamma \theta} -3u^2\,u_\alpha \right) + O(u^5), 
\\
C&=&C_0+ \frac{ a^2( b^2 -T) -T (b^2 -3T)  }{2DT^2( T - a^2)(b^2 - T) }u^2+O(u^4).
\end{eqnarray*}
For the actual numerical implementation, the equilibrium distribution
function can be calculated  analytically, up to any order of
accuracy required, by this procedure. Note that dependence on the temperature 
in the above equations is a nonperturbative result.

In order to establish the hydrodynamic equations corresponding to the present model,
 apart from the equilibrium pressure tensor
$P_{\alpha \beta}^{\rm eq}$ and the equilibrium third moment
$Q_{\alpha \beta \gamma }^{\rm eq}$, one needs to check also the fourth order moment,
$R_{\alpha\beta}^{\rm eq}=\sum_i c_{i\,\alpha}c_{i\,\beta} c^2f^{\rm eq}_i$. 
 A direct computation shows that the   equilibrium stress tensor is
  exact from the computational standpoint (accurate at least up to the
  order $O(u^8)$). The third moments $Q_{\alpha \beta \gamma}$ are
  accurate up to  the
order  $O(u\theta^2)$,
$O(u^3 \theta)$ and $O(u^5)$, where
$\theta = (T_0- T)/T_0$ is the deviation of the temperature from the
  reference value.  Finally, the 
fourth moments, $R_{\alpha \beta}$ are accurate  to the order 
$O(\theta^2)$, $O(u^2 \theta^2)$, $O(u^4)$.  Thus the
  Navier-Stokes-Fourier dynamics is recovered  to the order $O(u^3, \theta^2)$.
Notice  that if the temperature is fixed at the reference value $T=T_0$, the moment
$Q^{\rm eq}_{\alpha\beta\gamma}$ becomes exact  to the order $O(u^5)$, unlike in
the  second-order accurate standard lattice
Boltzmann models and  the isothermal model constructed above. 

   The local equilibrium
entropy, $S = - k_{\rm B} H_{\{w_i, \bc_i\}}(f^{\rm eq})$, for
the nonisothermal model satisfies the usual
expression for the entropy  of the ideal monoatomic gas to the overall
order of approximation of the method,
 $S= \rho\, k_{\rm B} \ln{\left(T^{D/2}/ \rho\right) } + O(u^4,
\theta^2)$.
Thus, the present model is able to retain the thermodynamics up to the
accuracy of the method. To the best of our knowledge, this is the
first discrete velocity model which is fully consistent with
thermodynamics.  

When the single relaxation time BGK model (\ref{LBM})  is used with
the present nonisothermal  equilibrium, the resulting transport coefficients
are as follows: For $D=1$, the  kinematic viscosity $\nu$ is equal to
zero, while the thermal conductivity $\kappa$ is  $\kappa = (3/2)(\tau
\,\rho)T$. For $D>1$, we have $\nu =(\tau \rho) T$,  and $
\kappa=((D+2)/2)(\tau \, \rho)T$ [we recall that one needs to renormalize the relaxation time in 
the BGK model as $\tau'=\tau \, \rho$, 
in order to obtain density-independent transport coefficients].

Finally, we present here some details of the  discretization scheme
 for model kinetic Eq. (\ref{LBM}). In a kinetic model there are two
 time scales one associated with  convection and the other one with
 collisions, where the timescale of  collisions is much smaller than
 that of convection. In order to have an efficient simulation
 scheme for the hydrodynamics, it is desirable to follow the time scale
 of the convection in the simulation. To achieve such a scheme we
 propose  to simulate discrete kinetic equation   
\begin{equation}
\label{lbe}
f_i(\bx, \delta  t) = {L}_{\{\bc_i, \delta t\}}\cdot
 \,
 \left[
f_i( \bx  , 0)  +
\frac{\alpha}{\left({2  \tau^\prime}+  \delta  t
\right)}
\left( f_i^{\rm eq}( \bx  , 0) - f_i( \bx  , 0)\right) \right],
\end{equation}
 were ${L}_{\{\bc_i, \delta t\}}$ is a linear convection operator and is defined by the relation         ${L}_{\{\bc_i, \delta t\}}\cdot g(\bx,t) = g(\bx -
          \bc_i \delta t, t)$, and the parameter $\alpha$ is  defined
by the condition:
\begin{equation}
\label{Eq:DHT}
H_{\{w_i, \bc_i\}}\left(f_i(\bx,0)\right) = H_{\{w_i, \bc_i\}}\left(f_i( \bx  , 0)  +
{\alpha}
\left( f_i^{\rm eq}( \bx  , 0) - f_i( \bx  , 0)\right)\right).
\end{equation}
Close to the local equilibrium the parameter $\alpha$ is equals to $2
\delta t$ \cite{AK2}. The details of the implementation of Eq. (\ref{Eq:DHT}) are
presented in the Ref. \cite{AK2}. The essence of the Eq. (\ref{lbe}) and (\ref{Eq:DHT}) is to separate the convection, which just
     shifts the population in space (operator ${L}_{\{\bc_i,
      \delta t\}}$), from the collisions  which represent the
    relaxation of the populations towards the equilibrium. After each
    convection step, the length of collision step is restricted by the condition
    of the entropy conservation during the collision (Eq.\ref{Eq:DHT}).     
This lumping of many short collision steps ensures that  rapid
     convergence towards hydrodynamics is achieved. The details
of the derivation of  Eq. (\ref{lbe}) and Eq. (\ref{Eq:DHT}) from the
     model kinetic equation (\ref{LBM}) will be presented elsewhere. 

In the isothermal case, this model can be implemented in a
 efficient manner by   taking   the uniform  grid spacing in each direction as 
$\delta x$ and  the time step as $\delta t = \delta x /
  \sqrt{3\,T_0}$. This means $\bx - \bc_i \delta  t $ is always a grid
 point and the resulting method is LBM.  In the nonisothermal  case,  the error in the approximations
   can be minimized by choosing the timestep as $\delta t = 3 \delta
 x/b$ and performing a dimensional splitting, for example in
   the two dimensional case ${L}_{\{c_{i x}, c_{i y}, \delta t\}}\cdot
   g(\bx,t) =  {L}_{\{c_{i x}, 0, \delta t\}}\cdot
   \left[  {L}_{\{0, c_{i y}, \delta t\}}\cdot
   g(\bx,t)\right]$. 

 For  directions in which
 velocity component  equals $a$, we propose to discretize the
 one dimensional convection operator by the Beam-Warming operator
 \cite{Laney} defined as $
{L}_{\{a, 3 \delta x/b  \}}\cdot g(x, t) = 
   0.02432 g(x, t) +0.99784 g(x- \delta x, t) + 0.02216 g(x- 2
   \delta x, t)$.
  We remind the reader that
 as the CFL number approaches the value of unity (in the present case
 equals $3 a/b = 0.95351$), the  error of the
 discretization vanishes \cite{Laney}.  The present discretization
 scheme, in contrast to the earlier proposed  finite-volume and finite-difference methods
 \cite{succi} for solving Eq. (\ref{LBM}), is using a large time   
step (of the order $O(\delta x)$) like LBM.
 
\begin{figure}[ht]
\begin{center}
\includegraphics[scale=0.35]{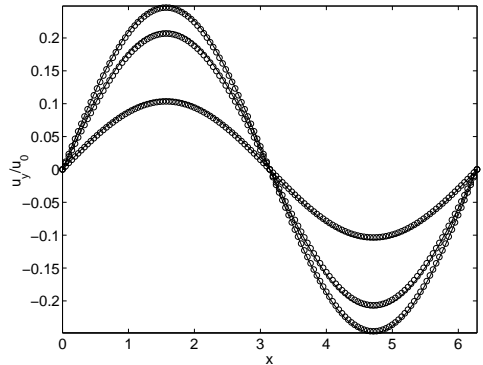}
\caption{\label{Fig1}
Simulation of the two dimension Taylor vortex flow. 
The velocity profiles at $y= \pi$ at three different times $t=0.03, 10,
50$ are shown. The solid lines
represent analytical results  with viscosity  $\nu= \tau^{\prime}T $, circles represent
simulation results.   
All the quantities are given in  dimensionless units.}
\end{center}
\end{figure}

   In order to check the effectiveness of the algorithm for the
   nonisothermal case, we have performed the simulation of the Taylor
   vortex flow. The flow is in the isothermal and  the incompressible
   set up, which is achieved in the simulation through the initial
   condition. This problem is chosen to validate the theoretical
   expression for the viscosity and the discretization procedure.  The flow is completely characterized by the analytical
   solution $u(x,y,t) =\bnabla \times \left[(u_0 / k_2) \exp{[-\nu
   (k_1^2+k_2^2) t]} \cos(k_1 x) \cos(k_2 y)\right]$. In the
   simulation, we have chosen $u_0= 0.0001, k_1 = 1, k_2 =4$. The result in Fig.\ref{Fig1} shows that the discretization procedure is
   working  well even at very short times.

To conclude, in this Letter we have derived minimal entropic kinetic models of the 
Boltzmann equation
for both isothermal and nonisothermal hydrodynamic simulations.
The resulting models have correct hydrodynamics,  they are
equipped with the appropriate $H$ function, and also they are  thermodynamically
consistent. A simple discretization scheme is proposed for the
simulation.  In the isothermal case, we have found analytically the corresponding 
local equilibrium in closed form (\ref{TED}), and thus proposed a
isothermal LBM with correct $H$ theorem. 
  
\bibliography{TBGK}

\end{document}